\documentclass[a4paper, orivec, runningheads]{llncs}
\usepackage{floatrow}
\newfloatcommand{capbtabbox}{table}[][\FBwidth]
\usepackage{tablefootnote}
\usepackage{graphicx}
\usepackage{amsmath}
\usepackage{array}
\usepackage{makecell, multirow, booktabs}
\usepackage{bm}
\usepackage{float}
\usepackage{url}
\usepackage{makeidx}  
\usepackage{graphicx}
\usepackage{subfigure}
\usepackage{amsmath}
\usepackage{amssymb}
\usepackage{verbatim}
\usepackage{algorithm}
\usepackage{algorithmic}
\usepackage{booktabs}
\usepackage{multirow}
\usepackage[misc]{ifsym} 
\usepackage{pbox} 
\usepackage{cite}
\usepackage{wrapfig}
\usepackage{threeparttable}
\usepackage{appendix}
\usepackage[colorlinks,
            linkcolor=blue,
            anchorcolor=blue,
            citecolor=blue
            ]{hyperref}
\usepackage{txfonts}
\begin{document}
\title{Scribble-Supervised Medical Image Segmentation via Dual-Branch Network and Dynamically Mixed Pseudo Labels Supervision}
\titlerunning{Scribble Supervised Segmentation via Pseudo Labels Supervision}
\author{Xiangde Luo\inst{1,2} \and Minhao Hu\inst{3} \and Wenjun Liao\inst{1} \and Shuwei Zhai\inst{1} \and Tao Song\inst{3} \and \\ Guotai Wang\inst{1,2}$^{(\textrm{\Letter})}$\and Shaoting Zhang\inst{1,2}}

\institute{
$^1$University of Electronic Science and Technology of China, Chengdu, China\\
$^2$Shanghai AI Lab, Shanghai, China\\
$^3$SenseTime Research, Shanghai, China\\
\url{guotai.wang@uestc.edu.cn}\\
\url{https://github.com/HiLab-git/WSL4MIS}\\
}
\maketitle

\begin{abstract}
Medical image segmentation plays an irreplaceable role in computer-assisted diagnosis, treatment planning and following-up. Collecting and annotating a large-scale dataset is crucial to training a powerful segmentation model, but producing high-quality segmentation masks is an expensive and time-consuming procedure. Recently, weakly-supervised learning that uses sparse annotations (points, scribbles, bounding boxes) for network training has achieved encouraging performance and shown the potential for annotation cost reduction. However, due to the limited supervision signal of sparse annotations, it is still challenging to employ them for networks training directly. In this work, we propose a simple yet efficient scribble-supervised image segmentation method and apply it to cardiac MRI segmentation. Specifically, we employ a dual-branch network with one encoder and two slightly different decoders for image segmentation and dynamically mix the two decoders' predictions to generate pseudo labels for auxiliary supervision. By combining the scribble supervision and auxiliary pseudo labels supervision, the dual-branch network can efficiently learn from scribble annotations end-to-end. Experiments on the public ACDC dataset show that our method performs better than current scribble-supervised segmentation methods and also outperforms several semi-supervised segmentation methods.\footnote{This is a comprehensive study about scribble-supervised medical image segmentation based on the \textcolor{red}{\textit{ACDC}} dataset and \textcolor{red}{\textit{WSL4MIS}} project.}
\keywords{Weakly-supervised learning \and scribble annotation \and pseudo labels}
\end{abstract}
\section{Introduction}
\begin{figure}[t]
  {\caption{Examples of dense and scribble annotations. BG, RV, Myo, LV, and UA represent the background, right ventricle, myocardium, left ventricle, and unannotated pixels respectively.
  }}
  {\includegraphics[width=1.0\linewidth]{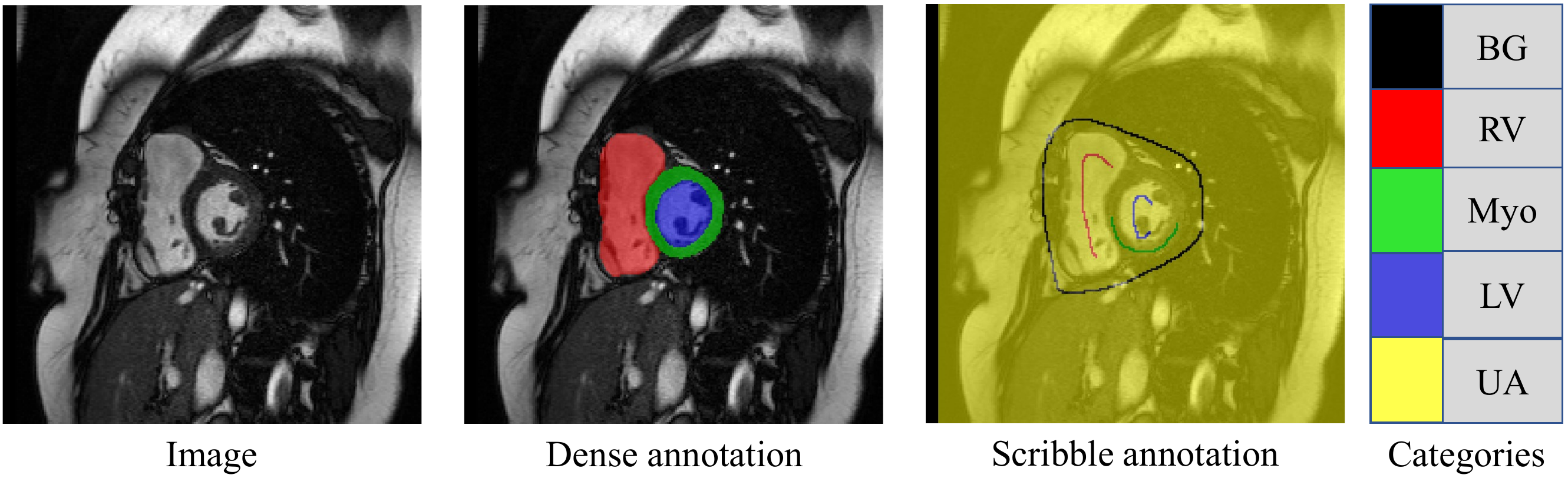}}
\end{figure}\label{diff}
Recently, Convolutional Neural Networks (CNNs) and Transformers have achieved encouraging results in automatic medical image segmentation~\cite{ronneberger2015u,isensee2021nnu,chen2021transunet}. Most of them need large-scale images with accurate pixel-level dense annotations to train models. However, collecting a large-scale and carefully annotated medical image dataset is still an expensive and time-consuming journey, as it requires domain knowledge and clinical experience~\cite{luo2020semi,luo2021mideepseg}. Recently, many efforts have been made to reduce the annotation cost for models training to alleviate this issue. For example, semi-supervised learning (SSL) combines a few labeled data and massive unlabeled data for network training~\cite{luo2020semi,luo2021urpc,bai2017semi}. Weakly supervised learning (WSL) uses sparse annotations to train models rather than dense annotations~\cite{dolz2021teach,valvano2021learning,dorent2021inter}. Considering collecting sparse annotations (points, scribbles and bounding boxes) is easier than dense annotations~\cite{lin2016scribblesup} and scribbles have better generality to annotate complex objects than bounding boxes and points~\cite{lin2016scribblesup,valvano2021learning} (Example in Fig.~\ref{diff}). This work focuses on exploring scribble annotations to train high-performance medical image segmentation networks efficiently and robustly.\\
\textit{\textbf{Scribble-Supervised Segmentation: }}Using scribble annotations to segment objects has been studied for many years. Before the deep learning era, combining user-provided sparse annotations and machine learning or other algorithms was the most popular and general segmentation method, such as GraphCuts~\cite{boykov2001interactive}, GrabCut~\cite{rother2004grabcut}, Random Walker~\cite{Grady2006}, GrowCut~\cite{Vezhnevets2005}, ITK-SNAP~\cite{yushkevich2006user}, Slic-Seg~\cite{Wang2016}, etc. Recently, deep learning with convolutional neural networks or transformers can learn to segment from dense annotations and then inference automatically. So, it is desirable to train powerful segmentation networks using scribble annotations. To achieve this goal, Lin et al \cite{lin2016scribblesup} proposed a graphical-based method to propagate information from scribbles to unannotated pixels and train models jointly. After that, Tang et al~\cite{tang2018regularized} introduced a Conditional Random Field (CRF) regularization loss to train segmentation networks directly. For medical images, Can et al~
\cite{can2018learning} proposed an iterative framework to train models with scribbles. At first, they seeded the scribbles into the Random Walker~\cite{Grady2006} to produce the initial segmentation. Then, they used the initial segmentation to train the model and refine the model's prediction with CRF for the network retraining. Finally, they repeated the second procedure several times for powerful segmentation models. Kim et al~\cite{kim2019mumford} proposed a level set-based~\cite{mumford1989optimal} regularization function to train deep networks with weak annotations. Lee et al~\cite{lee2020scribble2label} combined pseudo-labeling and label filtering to generate reliable labels for network training with scribble supervisions. Liu et al~\cite{liu2022weakly} presented a unified weakly-supervised framework to train networks from scribble annotations, which consists of an uncertainty-aware mean teacher and a transformation-consistent strategy. More recently, Valvano et al~\cite{valvano2021learning} proposed multi-scale adversarial attention gates to train models with mixed scribble and dense annotations. Although these attempts have saved the annotation cost by using scribble annotations, the performance is still lower than training with dense annotations, limiting the applicability in clinical practice.\\\textit{\textbf{Pseudo Labels for Segmentation: }}Pseudo labeling~\cite{lee2013pseudo} is widely used to generate supervision signals for unlabeled images/pixels. The main idea is utilizing imperfect annotations to produce high-quality and reliable pseudo labels for network training~\cite{chen2021semi,wang2021self}. Recently, some works have demonstrated~\cite{wang2021self,wu2021semi,luo2020eccv} that semi-supervised learning can benefit from high-quality pseudo labels. For weakly-supervised learning, Lee et al~\cite{lee2020scribble2label} showed that generating pseudo labels by ensembling predictions at a temporal level can boost performance. {\textit{Nevertheless, recent work~\cite{huo2021atso} points out the inherent weakness of these methods that the model retains the prediction from itself and thus resists updates.}} Recently, some works resort to perturbation-consistency strategy for semi-supervised learning~\cite{ouali2020semi,wu2021semi}, where the main branch is assisted by auxiliary branches that are typically perturbed and encouraged to produce similar predictions to the main branch. In this work, we assume that generating pseudo labels by mixing multiple predictions randomly can go against the above inherent weakness, as these auxiliary branches are added perturbations and do not enable interaction with each other.
\par Motivated by these observations, we present a simple yet efficient approach to learning from scribble annotations. Particularly, we employ a dual branches network (one encoder and two slightly different decoders) as the segmentation network. To learn from scribbles, the dual branches network is supervised by the partially cross-entropy loss ($pCE$), which only considers annotated pixels' gradient for back-propagation and ignores unlabeled pixels. At the same time, we employ the two predictions to generate hard pseudo labels for more substantial and more reliable supervision signals than scribbles. Afterward, we combine scribbles supervision and pseudo labels supervision to train the segmentation network end-to-end. Differently from threshold-based methods~\cite{lee2020scribble2label,wu2021semi}, we generate hard pseudo labels by dynamically mixing two branches' predictions, which can help against the inherent weakness~\cite{huo2021atso}. Such a strategy imposes the segmentation network to produce high-quality pseudo labels for unannotated pixels. We evaluate our method on a public scribble-supervised benchmark ACDC~\cite{bernard2018deep}. Experiments results show that our proposed method outperforms existing scribble-supervised methods when using the same scribble annotations and also performs better than semi-supervised methods when taking similar annotation budgets.

\begin{figure}[t]
  {\caption{Overview of the proposed method. The framework consists of an encoder ($\theta_e$), the main decoder ($\theta_{d1}$), and an auxiliary decoder ($\theta_{d2}$) and is trained with scribble annotations separately  ($L_{pCE}$). At the same time, the hard pseudo label is generated by dynamically mixing two decoders' outputs and used as the pseudo labels supervision for further network training ($L_{PLS}$).
  }}
  {\includegraphics[width=1.0\linewidth]{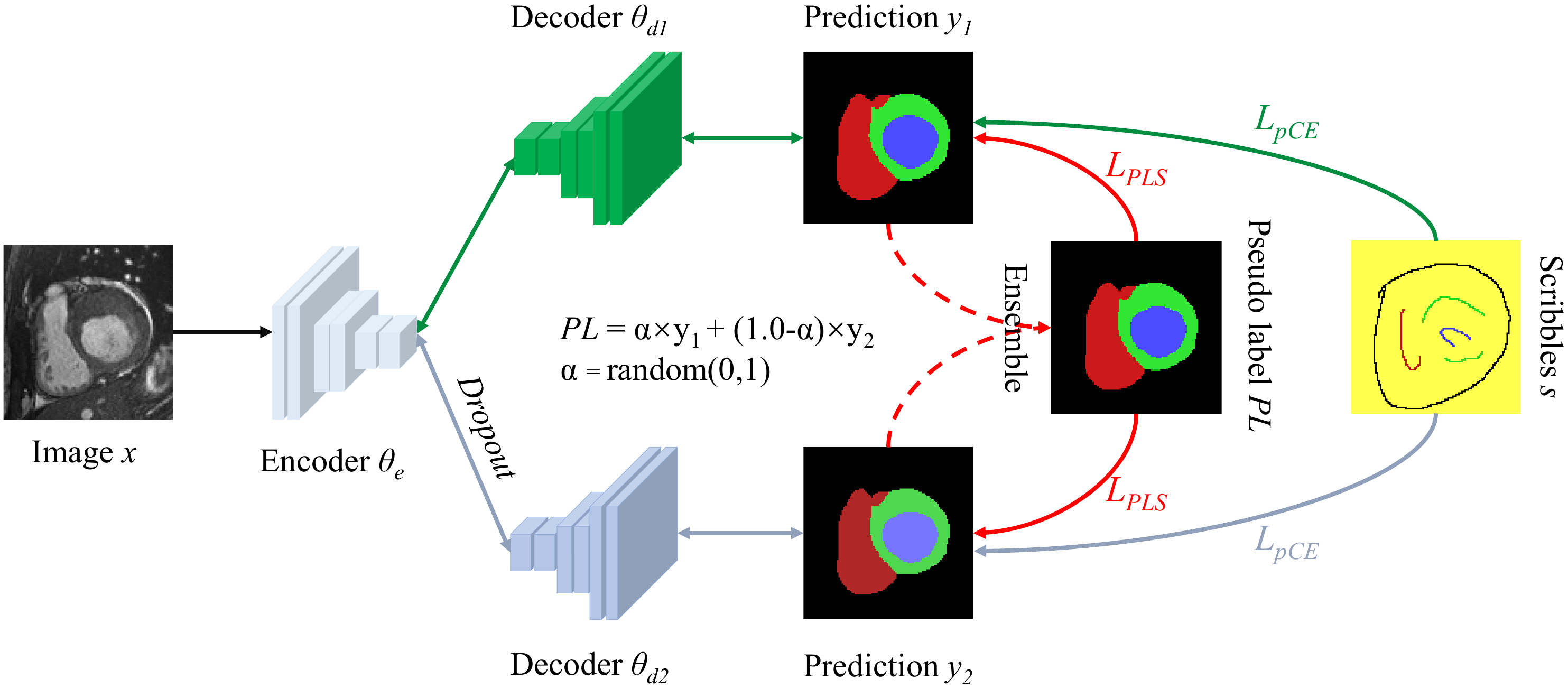}}
\end{figure}\label{pipeline}

\par The contributions of this work are two-fold. Firstly, we propose a dual-branch network and a dynamically mixed pseudo labeling strategy to train segmentation models with scribble annotations. Specifically, we generate high-quality hard pseudo labels by randomly mixing the two branches' outputs and use the generated pseudo labels to supervise the network training end-to-end. (2) Extensive experiments on the public cardiac MRI segmentation dataset (ACDC) demonstrate the effectiveness of the proposed method. Our method has achieved better performance on the ACDC dataset than existing scribble-supervised segmentation approaches and also outperformed several semi-supervised segmentation methods with similar annotation costs.
\section{Method}
The proposed framework for scribble-supervised medical image segmentation is depicted in Fig. 2. We firstly employ a network with one encoder and two slightly different decoders to learn from scribble annotations to segment target objects. At the same time, we utilize the two branches' outputs to generate hard pseudo labels that are used to assist the network training. Note that the training procedure is in an end-to-end manner rather than the multi-stage~\cite{can2018learning} or iterative refinement strategies~\cite{lin2016scribblesup}.
\subsection{Learning from Scribbles}For general scribble-supervised learning, the available dataset consists of images and scribble annotations , where the scribble is a set of pixels with a category or unknown label. Previous work~\cite{can2018learning} uses interactive segmentation methods~\cite{Grady2006} to propagate annotated pixels to the whole image for a rough segmentation and then train deep networks with the segmentation in a fully-supervised manner. Recently, there are much better alternatives~\cite{tang2018regularized,lee2020scribble2label}, e.g., using scribbles to train CNNs directly by minimizing a partial cross-entropy loss:
\begin{equation}
    L_{pCE}(y,s) = -\sum_c\sum_{i \in \omega_{s}}\log y_i^c
\end{equation}where $s$ represents the one-hot scribble annotations. $y_i^c$ is the predicted probability of pixel $i$ belonging class $c$. $\omega_{s}$ is the set of labeled pixels in $s$.
\subsection{Dual-branch Network}
The proposed network ($f(\theta_e, \theta_{d1}, \theta_{d2})$) is composed of a shared encoder ($\theta_e$) for feature extraction and two independent and different decoders ($\theta_{d1}$, $\theta_{d2}$) for segmentation and supplementary training (see Fig. 2). We embed a perturbed decoder into the general UNet~\cite{ronneberger2015u}, where the dropout~\cite{ouali2020semi} is used to introduce perturbation at the feature level. This design has two advantages: (1) It can be against the inherent weakness of pseudo-label in the single branch network~\cite{huo2021atso}, as the two branches' outputs are different due to the feature perturbation. (2) It can generate pseudo-label by two outputs ensemble but
does not require training two networks, and the encoder benefits from the two individual supervisions to boost the feature extraction ability~\cite{wu2021semi,ouali2020semi}. It is worthy to point out that some recent works used similar architecture for the consistency training~\cite{wu2021semi,ouali2020semi,luo2020semi} or knowledge distillation~\cite{dolz2021teach}. There are many significant differences in the learning scenarios and supervision strategies. Firstly, \cite{luo2020semi,wu2021semi,ouali2020semi} concentrate on semi-supervised learning and \cite{dolz2021teach} focus on knowledge distillation but we aim to scribble-supervised segmentation. Secondly, they employ consistency regularization to supervise networks, but we randomly mix two outputs to generate hard pseudo labels for fully supervised learning. These differences lead to different training, optimization strategies, and results.
\subsection{Dynamically Mixed Pseudo Labels}
Based on the dual-branch network, we further exploit the two decoders' outputs to boost the model training. We generate the hard pseudo labels by mixing two predictions dynamically, like mixup~\cite{zhang2017mixup}. The dynamically mixed pseudo labels (\textit{PL}) generation strategy is defined as:
\begin{equation}
    PL = argmax[\alpha \times y_{1} +(1.0-\alpha) \times y_{2}],~\alpha = random(0, 1)
\end{equation}where $y_{1}$ and $y_{2}$ are outputs of decoder 1 and 2, respectively. $\alpha$ is randomly generated in (0, 1) at each iteration. This strategy boosts the diversity of pseudo labels and avoids the inherent weakness of the pseudo labeling strategy (remembering itself predictions without updating )~\cite{huo2021atso}. $argmax$ is used to generate hard pseudo labels. Compared with consistency learning~\cite{ouali2020semi,wu2021semi}, this strategy cuts off the gradient between $\theta_{d1}$ and $\theta_{d2}$ to maintain their independence rather than enforce consistency directly. In this way, the supervision signal is enlarged from a few pixels to the whole image, as the scribbled pixels are propagated to all unlabeled pixels by the dynamically mixed pseudo labeling. Then, we further employ the generated $PL$ to supervise $\theta_{d1}$ and $\theta_{d2}$ separately to assist the network training. The \textit{P}seudo \textit{L}abels \textit{S}upervision (\textit{PLS}) is defined as :
\begin{equation}
    L_{PLS}(PL, y_1, y_2) = 0.5 \times (L_{Dice}(PL, y_1) + L_{Dice}(PL, y_2))
\end{equation}where $L_{Dice}$ is the widely-used dice loss and also can be replaced by cross-entropy loss or other segmentation loss functions. Finally, the proposed network can be trained with scribble annotations by minimizing the following joint object function:
\begin{equation}\label{equ:total_loss}
    L_{total} = \underbrace{0.5 \times (L_{pCE}(y_1,s) + L_{pCE}(y_2,s))}_{scribble~supervision} + \lambda \times \underbrace{L_{PLS}(PL, y_1, y_2)}_{pseudo~labels~supervision}
\end{equation}$\lambda$ is a weight factor to balance the supervision of scribbles and pseudo labels.

\section{Experiment and Results}
\subsection{Experimental Details}\textit{\textbf{Dataset: }}We evaluate the proposed method on the training set of ACDC~\cite{bernard2018deep} via five-fold cross-validation. This dataset is publicly available, with 200 short-axis cine-MRI scans from 100 patients, and each patient has two annotated end-diastolic (ED), and end-systolic (ES) phases scans. And each scan has three structures' dense annotation, including the right ventricle (RV), myocardium (Myo), and left ventricle (LV). Recently, Valvano et al~\cite{valvano2021learning} provided the scribble annotation for each scan manually. Following previous works~\cite{bai2017semi,valvano2021learning}, we employ the 2D slice segmentation rather than 3D volume segmentation, as the thickness is too large.\\\textit{\textbf{Implementation Details: }}
We employed the UNet~\cite{ronneberger2015u} as the base segmentation network architecture, and we further extended the basic UNet to dual branches network by embedding an auxiliary decoder. We added the dropout layer (ratio=0.5) before each conv-block of the auxiliary decoder to introduce perturbations. We implemented and ran our proposed and other comparison methods by PyTorch~\cite{paszke2019pytorch} on a cluster with 8 TiTAN 1080TI GPUs. For the network training, we first re-scaled the intensity of each slice to 0-1. Then, random rotation, random flipping, random noise were used to enlarge the training set, and the augmented image was resized to 256 $\times$ 256 as the network input. We used the SGD (weight decay = $10^{-4}$, momentum = 0.9) to minimize the joint object function Eq.~\ref{equ:total_loss} for the model optimization. The poly learning rate strategy was used to adjust the learning rate online~\cite{luo2021urpc}. The batch size, total iterations, and $\lambda$ are set to 12, $60k$, and 0.5, respectively. For testing, we produced predictions slice by slice and stacked them into a 3D volume. For a fair comparison, we used the primary decoder's output as the final result during the inference stage and did not use any post-processing method. Note that all experiments were conducted in the same experimental setting. The 3D Dice Coefficient ($DSC$) and 95\% Hausdorff Distance ($HD_{95}$) are used as evaluation metrics. All code, data, and details of existing and proposed methods at:~\url{https://github.com/HiLab-git/WSL4MIS}.
\begin{table}[t]
\centering
\scriptsize
\renewcommand\arraystretch{0.4}
\caption{Comparison with existing weakly-/semi-supervised methods on the ACDC dataset. All results are based on the \textit{5-fold cross-validation} with same backbone (UNet). Mean and standard variance (in parentheses) values of 3D \textit{DSC} and \textit{$HD_{95}$} (mm) are presented in this table. \textbf{$^*$} denotes p-value $<$ 0.05 (paired t-test) when comparing with the second place method (RLoss~\cite{tang2018regularized}).}
\setlength{\tabcolsep}{0.25mm}{
\begin{tabular}{rlllllllll}
\hline
\multirow{2}{*}{Type}&\multirow{2}{*}{Method}&\multicolumn{2}{c}{\textbf{RV}}&\multicolumn{2}{c}{\textbf{Myo}}&\multicolumn{2}{c}{\textbf{LV}}&\multicolumn{2}{c}{\textbf{Mean}} \\\cline{3-10} 
&&\textit{DSC}&\textit{$HD_{95}$}&\textit{DSC}&\textit{$HD_{95}$}&\textit{DSC}&\textit{$HD_{95}$}&\textit{DSC}&\textit{$HD_{95}$}\\
\hline
&pCE\cite{lin2016scribblesup}&0.625(0.16)&
187.2(35.2)&
0.668(0.095)&
165.1(34.4)&
0.766(0.156)&
167.7(55.0)&
0.686(0.137)&
173.3(41.5)\\
&RW\cite{Grady2006}&0.813(0.113)&
11.1(17.3)&
0.708(0.066)&
9.8(8.9)&
0.844(0.091)&
9.2(13.0)&
0.788(0.09)&
10.0(13.1)\\
&USTM\cite{liu2022weakly}&0.815(0.115)&
54.7(65.7)&
0.756(0.081)&
112.2(54.1)&
0.785(0.162)&
139.6(57.7)&
0.786(0.119)&
102.2(59.2 )\\
\multirow{2}{*}{WSL}&S2L\cite{lee2020scribble2label}&0.833(0.103)&
14.6(30.9)&
0.806(0.069)&
37.1(49.4)&
0.856(0.121)&
65.2(65.1)&
0.832(0.098)&
38.9(48.5)\\
&MLoss\cite{kim2019mumford}&0.809(0.093)&
17.1(30.8)&
0.832(0.055)&
28.2(43.2)&
0.876(0.093)&
37.9(59.6)&
0.839(0.080)&
27.7(44.5)\\
&EM\cite{grandvalet2005semi}&0.839(0.108)&
25.7(44.5)&
0.812(0.062)&
47.4(50.6)&
0.887(0.099)&
43.8(57.6)&
0.846(0.089)&
39.0(50.9)\\
&RLoss\cite{tang2018regularized}&0.856(0.101)&
7.9(12.6)&
0.817(0.054)&
\textbf{6.0(6.9)}&
0.896(0.086)&
\textbf{7.0(13.5)}&
0.856(0.080)&
\textbf{6.9(11.0)}\\
&\textbf{Ours}&\textbf{0.861(0.096)}&
\textbf{7.9(12.5)}&
\textbf{0.842(0.054)}$^\textbf{*}$&
9.7(23.2)&
\textbf{0.913(0.082)}$^\textbf{*}$&
12.1(27.2)&
\textbf{0.872(0.077)}$^\textbf{*}$&9.9(21.0)\\
\hline
&PS\cite{ronneberger2015u}&0.659(0.261)&
26.8(30.4)&
0.724(0.176)&
16.0(21.6)&
0.790(0.205)&
24.5(30.4)&
0.724(0.214)&
22.5(27.5)\\
\multirow{2}{*}{SSL}&DAN\cite{zheng2019semi}&0.639(0.26)&
20.6(21.4)&
0.764(0.144)&
9.4(12.4)&
0.825(0.186)&
15.9(20.8)&
0.743(0.197)&
15.3(18.2)\\
&AdvEnt\cite{vu2019advent}&0.615(0.296)&
20.2(19.4)&
0.760(0.151)&
8.5(8.3)&
0.848(0.159)&
11.7(18.1)&
0.741(0.202)&
13.5(15.3)\\
&MT\cite{tarvainen2017mean}&0.653(0.271)&
18.6(22.0)&
0.785(0.118)&
11.4(17.0)&
0.846(0.153)&
19.0(26.7)&
0.761(0.180)&
16.3(21.9)\\
&UAMT\cite{yu2019uncertainty}&0.660(0.267)&
22.3(22.9)&
0.773(0.129)&
10.3(14.8)&
0.847(0.157)&
17.1(23.9)&
0.760(0.185)&
16.6(20.5)\\
\hline
{FSL}&FullSup\cite{ronneberger2015u}&0.882(0.095)&
6.9(10.8)&
0.883(0.042)&
5.9(15.2)&
0.930(0.074)&
8.1(20.9)&
0.898(0.070)&
7.0(15.6)\\

\hline
\end{tabular}}
\label{tab:sota_acdc}
\end{table}

\subsection{Results}
\textit{\textbf{Comparison with Other Methods: }}Firstly, we compared our method with seven scribble-supervised segmentation methods with the same set of scribbles: 1) pCE only~\cite{lin2016scribblesup} (lower bound), 2) using pxeudo label generated by Random Walker (RW)~\cite{Grady2006}, 3) Uncertainty-aware Self-ensembling and Transformation-consistent Model (USTM)~\cite{liu2022weakly}, 4) Scribble2Label (S2L)~\cite{lee2020scribble2label}, 5) Mumford–shah Loss (MLoss)~\cite{kim2019mumford}, 6) Entropy Minimization (EM)~\cite{grandvalet2005semi}, 7) Regularized Loss (RLoss)~\cite{tang2018regularized}. The first section of Table~\ref{tab:sota_acdc} lists the quantitative comparison of the proposed with seven existing weakly supervised learning methods. It can be found that our method achieved the best performance in terms of mean \textit{DSC} (p-value $<$ 0.05) and second place in the \textit{$HD_{95}$} metric than other methods. \\ Afterward, we further compared our method with other popular annotation-efficient segmentation methods, e.g., semi-supervised learning methods. Following~\cite{dorent2021inter}, we investigated the performance difference of these approaches when using very similar annotation costs. To do so, we trained networks with partially supervised and semi-supervised fashions, respectively. We used a 10\% training set (8 patients) as labeled data and the remaining as unlabeled data, as the scribble annotation also takes similar annotation costs~\cite{valvano2021learning}. For partially supervised (\textit{PS}) learning, we used the 10\% labeled data to train networks only. For semi-supervised learning, we combined the 10\% labeled data and 90\% unlabeled data to train models jointly. We further employed four widely-used semi-supervised segmentation methods for comparison: 1) Deep Adversarial Network (DAN)~\cite{zheng2019semi}, 2) Adversarial Entropy Minimization (AdvEnt)~\cite{vu2019advent}, 3) Mean Teacher (MT)~\cite{tarvainen2017mean}, and Uncertainty Aware Mean Teacher (UAMT)~\cite{yu2019uncertainty}. The quantitative comparison is presented in the second section of Table~\ref{tab:sota_acdc}. It shows that the scribbled annotation can achieve better results than pixel-wise annotation when taking a similar annotation budget. Moreover, our weakly-supervised method significantly outperforms existing semi-supervised methods in the cardiac MR segmentation. Finally, we also investigated the upper bound when using all mask annotation to train models (FullSup) in the last row of Table~\ref{tab:sota_acdc}. It can be found that our method is slightly inferior compared with fully supervised learning with pixel-wise annotation. But our method requires fewer annotation costs than pixel-wise annotation. Fig.~3 shows the segmentation results obtained by existing and our methods, and the corresponding ground truth on the ACDC dataset (patient026\_frame01). We can observe that the result obtained by our method is more similar to the ground truth than the others. It further shows that drawing scribble is a potential data annotation approach to reduce annotation costs.\\\textit{\textbf{Sensitivity analysis of $\lambda$: }}The study was conducted to assess the sensitivity of $\lambda$ in Eq.~\ref{equ:total_loss}. Particularly, the \textit{PLS} term plays a crucial role in the proposed framework, as it controls the usage of the pseudo labels during the network training. We investigated the segmentation performance of the proposed framework when the $\lambda$ is set to $\{0.01, 0.1, 0.2, 0.3, 0.5, 1.0\}$. Fig.~4 shows the evolution of the segmentation result of RV, Myo, LV, and their average results, all these results are based on the 5-fold cross-validation. It can be observed that increasing $\lambda$ from 0.01 to 0.5 leads to better performance in terms of both \textit{DSC} and \textit{$HD_{95}$}. When the $\lambda$ is set to 1.0, the segmentation result just decreases slightly compared with 0.5 (0.872 vs 0.870 in term of mean \textit{DSC}). These observations show that the proposed method is not sensitive to $\lambda$.\\
\begin{figure}[t]
  {\caption{Qualitative comparison of our proposed method and several existing ways.
  }}
  {\includegraphics[width=1.0\linewidth]{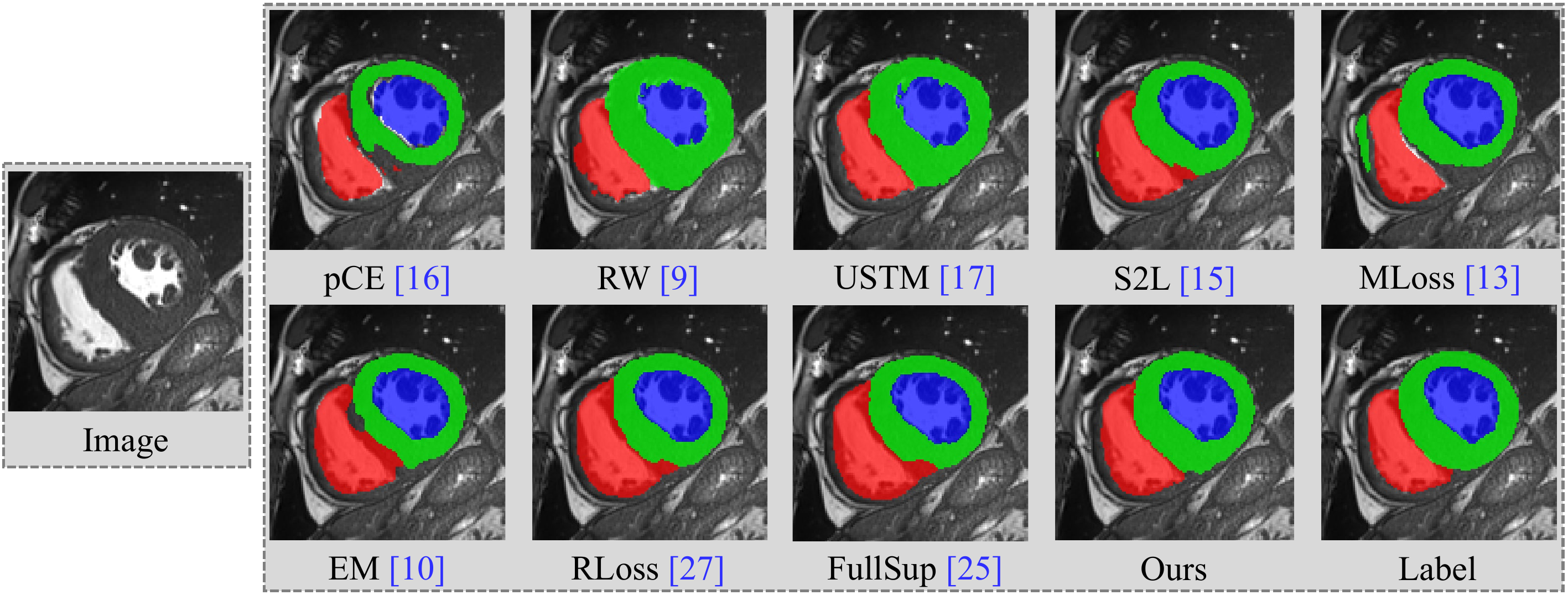}}
\end{figure}\label{visual}
\begin{figure}[t]
  {\caption{Sensitivity analysis of hyper-parameter $\lambda$.
  }}
{\includegraphics[width=1.0\linewidth]{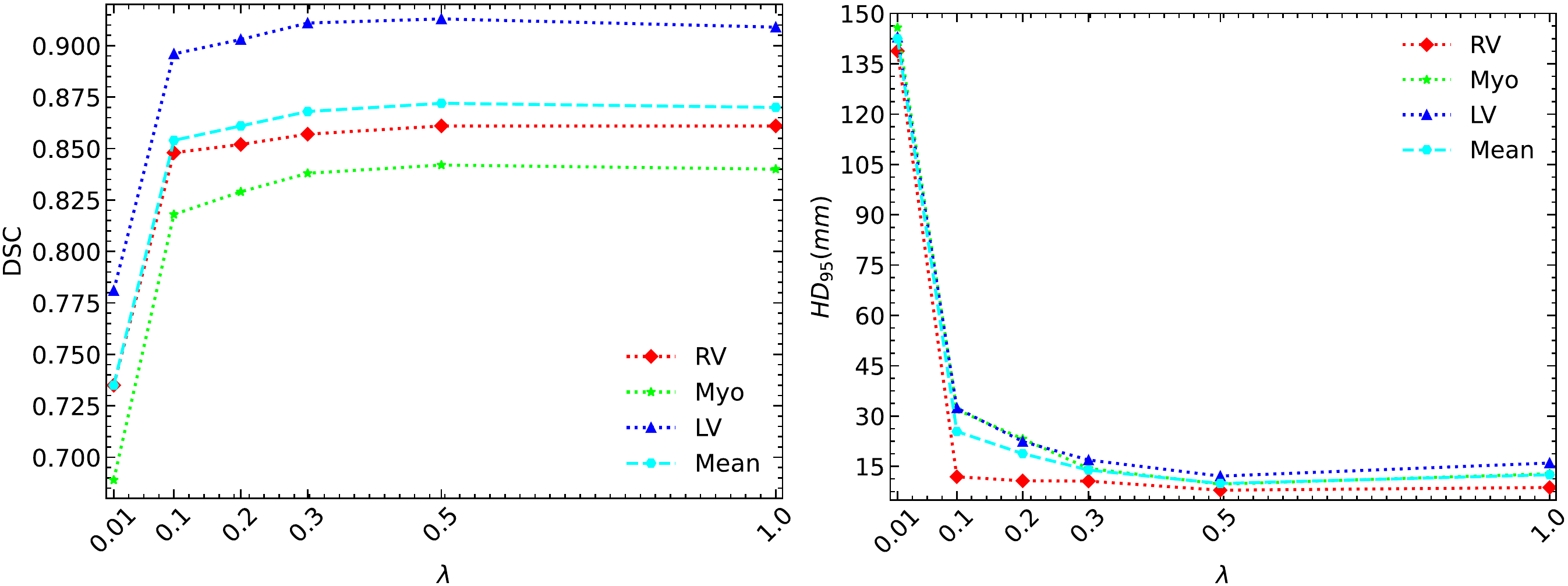}}
\end{figure}\label{abla_study}\textit{\textbf{Ablation Study: }}We further investigated the effect of using different supervision approaches for the dual-branch network: 1) Consistency Regularization (\textit{CR})~\cite{dolz2021teach} that encourages the two predictions to be similar, directly; 2) Cross Pseudo Supervision (\textit{CPS})~\cite{wu2021semi,chen2021semi} that uses one decoder's output as the pseudo label to supervise the other one; 3) the proposed approach dynamically mixes two outputs to generate hard pseudo labels for two decoders training separately. We trained the dual-branch network with scribbles and the above supervision strategies. The quantitative evaluation results are presented in Table~\ref{tab:abla_conpent}. It can be observed that compared with \textit{CR} and \textit{CPS}, using our proposed \textit{PLS} leads to the best performance. Moreover, we also investigated the performance when $\alpha$ is set to a fixed value (0.5) and dynamic values. The result demonstrates the effectiveness of the proposed dynamically mixing strategy. In addition, we found that the main ($\theta_{d1}$) and auxiliary ($\theta_{d2}$) decoders achieve very similar results.

\begin{table}[t]
\centering
\scriptsize
\renewcommand\arraystretch{0.4}
\caption{Ablation study on different supervision strategies for the dual-branch network. Single denotes the baseline UNet~\cite{ronneberger2015u} with \textit{pCE} only. \textit{CR} means consistency regularization between the main and auxiliary decoders~\cite{dolz2021teach}. \textit{CPS} is the cross pseudo supervision strategy in \cite{wu2021semi,chen2021semi}. \textit{Ours} is  proposed \textit{PLS}, $\theta_{d1}$ and $\theta_{d2}$ mean the prediction of main and auxiliary decoders, respectively.}
\setlength{\tabcolsep}{0.25mm}{
\begin{tabular}{lllllllll}
\hline
\multirow{2}{*}{Method}&\multicolumn{2}{c}{\textbf{RV}}&\multicolumn{2}{c}{\textbf{Myo}}&\multicolumn{2}{c}{\textbf{LV}}&\multicolumn{2}{c}{\textbf{Mean}} \\\cline{2-9} 
&\textit{DSC}&\textit{$HD_{95}$}&\textit{DSC}&\textit{$HD_{95}$}&\textit{DSC}&\textit{$HD_{95}$}&\textit{DSC}&\textit{$HD_{95}$}\\
\hline
Single\cite{lin2016scribblesup}&0.625(0.16)&
187.2(35.2)&
0.668(0.095)&
165.1(34.4)&
0.766(0.156)&
167.7(55.0)&
0.686(0.137)&
173.3(41.5)\\
\hline
Dual+\textbf{\textit{CR}\cite{dolz2021teach}}&0.844(0.106)&
20.1(37.2)&
0.798(0.07)&
62.2(55.7)&
0.873(0.101)&
63.4(65.5)&
0.838(0.092)&
48.6(52.8)\\
Dual+\textbf{\textit{CPS}\cite{wu2021semi,chen2021semi}}& 0.849(0.099)&
12.4(25.6)&
0.833(0.056)&
19.3(33.5)&
0.905(0.091)&
18.3(35.8)&
0.863(0.082)&
16.6(31.6)\\
\hline
\textbf{\textit{Ours}} ($\alpha$=0.5, $\theta_{d1}$)&0.855(0.101)&8.6( 13.9 ) &
0.837(0.053)&13.6(29.1)&0.908(0.086)&15.8(34.1)&0.866(0.08)&12.6(25.7)\\
\textbf{\textit{Ours}} ($\alpha$=random, $\theta_{d1}$)&\textbf{0.861(0.096)}&
7.9(12.5)&
\textbf{0.842(0.054)}&
\textbf{9.7(23.2)}&
\textbf{0.913(0.082)}&
12.1(27.2)&
\textbf{0.872(0.077)}&9.9(21.0)\\
\textbf{\textit{Ours}} ($\alpha$=random, $\theta_{d2}$)&0.861(0.098)&
\textbf{7.3(10.3)}&
0.840(0.058)&
10.9(24.5)&
0.911(0.086)&
\textbf{11.3(26.4)}&
0.871(0.08)&
\textbf{9.8(20.4)}\\
\hline
\end{tabular}}
\label{tab:abla_conpent}
\end{table}

\section{Conclusion}
In this paper, we presented pseudo labels supervision strategy for scribble-supervised medical image segmentation. A dual-branch network is employed to learn from scribble annotations in an end-to-end manner. Based on the dual-branch network, a dynamically mixed pseudo labeling strategy was presented to propagate the scribble annotations to the whole image and supervise the network training. Experiments on a public cardiac MR image segmentation dataset (ACDC) demonstrated the effectiveness of the proposed method, where it outperformed seven recent scribble-supervised segmentation methods using the same scribble annotations and four semi-supervised segmentation methods with very similar annotation costs. In the future, we will extend and evaluate the proposed method on other challenging medical image segmentation tasks.
\bibliographystyle{splncs04}
\bibliography{ref.bib}

\end{document}